\documentclass[11pt,conference,letter]{IEEEtran}
\usepackage{times}
\usepackage{epsfig}
\usepackage{amsmath}
\usepackage{amsfonts}
\usepackage{amssymb}
\usepackage{amstext}
\usepackage{latexsym}
\usepackage{color}
\usepackage{ifthen}
\usepackage{multirow}
\usepackage{subfigure}
\newcommand{\be}{\begin{equation}}
\newcommand{\ee}{\end{equation}}
\newcommand{\bear}{\begin{eqnarray}}
\newcommand{\eear}{\end{eqnarray}}
\newcommand{\bears}{\begin{eqnarray*}}
\newcommand{\eears}{\end{eqnarray*}}
\newcommand{\bi}{\begin{itemize}}
\newcommand{\ei}{\end{itemize}}
\newcommand{\ben}{\begin{enumerate}}
\newcommand{\een}{\end{enumerate}}

\newcommand{\beq}{\begin{equation}}
\newcommand{\eeq}{\end{equation}}

\newtheorem{theorem}{Theorem}[section]

\newtheorem{defn}[theorem]{Definition} 
\newtheorem{lemma}[theorem]{Lemma}
\newtheorem{corollary}[theorem]{Corollary}

\newcommand{\ubf}{\mbox{${\bf u }$} }

\newcommand{\xbf}{\mbox{${\bf x }$} }
\newcommand{\ybf}{\mbox{${\bf y }$} }
\newcommand{\zbf}{\mbox{${\bf z }$} }

\newcommand{\Fbf}{\mbox{${\bf F }$} }
\newcommand{\Gbf}{\mbox{${\bf G }$} }

\newcommand{\Ibf}{\mbox{${\bf I }$} }

\newcommand{\Xbf}{\mbox{${\bf X }$} }

\newcommand{\Zerobf}{\mbox{${\large{\bf 0 }}$} }

\newcommand{\FF}{\mbox{$\mathbb{F}$} }

\newcommand{\Prob}{\mbox{${\mathbb P}$} }

\newcommand{\prob}[1]{\Prob \left\{ #1 \right\}}

\begin{document}

\title{Wireless Network Information Flow}

\author{\authorblockN{Amir Salman Avestimehr}
\authorblockA{Wireless Foundations\\
 UC Berkeley,\\
 Berkeley, California, USA.\\
Email: {\sffamily avestime@eecs.berkeley.edu}} \and
\authorblockN{Suhas
N. Diggavi}
\authorblockA{School of Computer and\\
Communication Sciences, EPFL,\\
 Lausanne, Switzerland.\\
Email: {\sffamily suhas.diggavi@epfl.ch}}
\and
\authorblockN{David N C. Tse}
\authorblockA{ Wireless Foundations\\
 UC Berkeley,\\
 Berkeley, California, USA.\\
Email: {\sffamily dtse@eecs.berkeley.edu}}
 }

\maketitle

\begin{abstract}

We present an achievable rate for general deterministic relay
networks, with broadcasting at the transmitters and interference at
the receivers. In particular we show that if the optimizing
distribution for the information-theoretic cut-set bound is a product
distribution, then we have a complete characterization of the
achievable rates for such networks. For linear deterministic finite-field
models discussed in a companion paper \cite{ADTallerton1_07}, this is indeed the case, and we have a generalization of
the celebrated max-flow min-cut theorem for such a network.

\end{abstract}

%First part of this paper includes:
%1-Introduction
%2-Deterministic Wireless Network Model
%3-Single-Source, Single Destination Network and its Capacity

\section{Introduction}
\label{sec:intro}

Consider a network represented by a directed relay network
$\mathcal{G}=(\mathcal{V},\mathcal{E})$ where $\mathcal{V}$ are the
vertices representing the communication nodes in the relay network.
The communication problem considered is unicast (or multicast with all
destinations requesting the {\em same} message). Therefore a special
node $S\in\mathcal{V}$ is considered the source of the message and a
special node $D\in\mathcal{V}$ is the intended destination. All other
nodes in the network facilitate communication between $S$ and $D$.  In
a wireline network, such as studied in \cite{ACLY00}, the edges
$\mathcal{E}$ of the network do not interact and are orthogonal
communication channels.  In this paper, transmissions are
{\em not necessarily} orthogonal and signals sent by the nodes in
$\mathcal{V}$ can in general broadcast and also interfere with one another. In
particular, for each vertex $j\in\mathcal{V}$ of the network, there is
only one transmitted signal $x_j$ which is broadcast to the other
nodes connected to this vertex. Moreover it has only one received
signal $y_j$ which is a deterministic function of all the signals
transmitted by the nodes connected to it. By connection we mean the
nodes that have edges belonging to the set $\mathcal{E}$. By
deterministic we mean that $y_j = g_j(\{x_k\}_{k\in\mathcal{N}_j})$,
where $\mathcal{N}_j$ is the input neighbors of node $j$. Therefore,
we have deterministic broadcast and multiple access channels
incorporated into the model to reflect physical layer effects.

This approach is motivated by the development of the linear
deterministic finite-field model for wireless channels
\cite{ADTitw07}, and its connection to Gaussian relay networks
\cite{ADTallerton1_07}.  Historically, deterministic relay networks
were perhaps first studied in \cite{Aref_thesis}, where a
deterministic model with broadcast but {\em no multiple access} was
studied (the so-called Aref's networks).  For such a network, the
unicast capacity was determined in \cite{Aref_thesis} and its
extension to multicast capacity when all receivers needed the same
message was done in \cite{RK06}. A three-node deterministic relay
network capacity was characterized in \cite{AE82}, where {\em both}
broadcast and multiple access were allowed. Network coding is
information flow on a very special class of deterministic networks,
where all the links are non-interfering and orthogonal. For such
networks, the unicast capacity is given by the classical max-flow
min-cut theorem of Ford-Fulkerson, and the multicast capacity has been
determined in the seminal work \cite{ACLY00}.  More recently, the
capacity of a class of erasure relay networks has been established
where random erasures attempt to model the noise and collisions
\cite{GBS06}.  In all these cases, where the characterization exists,
the information-theoretic cut-set was achievable. Recently, a relay
network where the cut-set bound is not tight has been demonstrated in
\cite{YuISIT07}.

We first consider general deterministic functions to model the
broadcast and multiple access channels. For such networks we show an
achievability which is tight only for functions and networks where the
independent input distribution optimizes the information-theoretic
cut-set bound. For Aref's networks where there is no interference,
this is indeed the case and our result is a generalization of his. For
deterministic networks where there {\em is} interference but the
deterministic functions are linear over a finite field, it turns out
that the cut-set bound is also optimized by the product
distribution. For this case, our result is a natural generalization of
the celebrated max-flow min-cut theorem.  These ideas are easily
extended to the multicast case, where we want to simultaneously
transmit one message from $S$ to all destinations $D$ in the set
$\mathcal{D}$. For the linear finite-field model, we characterize
the multicast capacity, and therefore generalize the result in
\cite{ACLY00}.  We will discuss this in more detail in the next
section.

%%This paper is organized as follows. In Section \ref{sec:MainRes}, we state the main results %%of the paper and their interpretations. We give an illustration of proof ideas in section %%\ref{secPfIdea}. Then we we focus on networks that have a layered structure and we prove the %%results for these networks. To make the paper more readable the result for layered networks %%are proved first for linear deterministic model that is more intuitive in section %%\ref{sec:LayFF} and then for general deterministic model in section \ref{sec:GenDet}. We %%then use the idea of time-expanded representation of the network and submodularity %%properties of entropy to finish the proofs for general networks in section %%\ref{sec:TimExp}. We conclude with a discussion of simple extensions and open
%%questions in Section \ref{sec:Disc}.

\section{Problem statement and main results}
\label{sec:MainRes}

\subsection{General Deterministic network}
\label{subsec:MainResGenDet}
As stated in Section \ref{sec:intro}, we consider a directed network
$\mathcal{G}=(\mathcal{V},\mathcal{E})$, where the received signal
$y_j$ at node $j\in\mathcal{V}$ is given by
\begin{equation}
\label{eq:GenDetModel}
y_j = g_j(\{x_i\}_{i\in\mathcal{N}_j}),
\end{equation}
where we define the input neighbors $\mathcal{N}_j$ of $j$ as the set
of nodes whose transmissions affect $j$, and can be formally defined
as $\mathcal{N}_j = \{i: (i,j)\in\mathcal{E}\}$. Note that this
implies a deterministic multiple access channel for node $j$ and a
deterministic broadcast channel for the transmitting nodes.

For any relay network, there is a natural information-theoretic
cut-set bound \cite{CoverThomas91}, which upperbounds the reliable
transmission rate $R$. Applied to our model, we have:
\begin{eqnarray}
\nonumber
R &< &\max_{p(\{x_j\}_{j\in\mathcal{V}})} \min_{\Omega\in\Lambda_D}
I(Y_{\Omega^c};X_{\Omega}|X_{\Omega^c}) \\ \label{eq:CutSet} & \stackrel{(a)}{=} &
\max_{p(\{x_j\}_{j\in\mathcal{V}})} \min_{\Omega\in\Lambda_D}
H(Y_{\Omega^c}|X_{\Omega^c})
\end{eqnarray}
where $\Lambda_D=\{\Omega:S\in\Omega,D\in\Omega^c\}$ is all
source-destination cuts (partitions) and $(a)$ follows since we are
dealing with deterministic networks.

The following are our main results for general deterministic
networks.
\begin{theorem}
\label{thm:GenDetNet}
Given a general deterministic relay network (with broadcast and
multiple access), we can achieve all rates  $R$ up to,
\begin{equation}
\label{eq:GenDetNet}
\max_{\prod_{i\in\mathcal{V}} p(x_i)} \min_{\Omega\in\Lambda_D}
H(Y_{\Omega^c}|X_{\Omega^c})
\end{equation}
\end{theorem}
This theorem easily extended to the multicast case, where we want to simultaneously transmit one
message from $S$ to all destinations in the set $D\in\mathcal{D}$:
\begin{theorem}
\label{thm:GenDetNetMulticast}
Given a general deterministic relay network (with broadcast and
multiple access), we can achieve all rates $R$ from $S$ multicasting to all destinations $D\in\mathcal{D}$
up to,
\begin{equation}
\label{eq:GenDetNetMulticast}
\max_{\prod_{i\in\mathcal{V}} p(x_i)} \min_{D\in\mathcal{D}} \min_{\Omega\in\Lambda_D}
H(Y_{\Omega^c}|X_{\Omega^c})
\end{equation}
\end{theorem}

This achievability result in Theorem \ref{thm:GenDetNet} extends the
results in \cite{RK06} where only deterministic broadcast network
(with no interference) were considered.

Note that when we compare (\ref{eq:GenDetNet}) to the cut-set upper bound in (\ref{eq:CutSet}),
we see that the difference is in the maximizing set {\em i.e.,} we are
only able to achieve independent (product) distributions whereas the
cut-set optimization is over any arbitrary distribution. In
particular, if the network and the deterministic functions are such
that the cut-set is optimized by the product distribution, then we
would have matching upper and lower bounds. This indeed happens when
we consider the linear finite-field model discussed below.

\subsection{Linear Finite-Field Deterministic network}
\label{subsec:MainResLinDet}

A special deterministic model which is motivated
\cite{ADTallerton1_07} by its close connection to the Gaussian
model is the linear finite-field model, where the received signal
$\ybf_j\in\FF_p^q$ is a vector defined over a finite field $\FF_p$
given by,
\begin{equation}
\label{eq:LinDetModel}
\ybf_{j}=\sum_{i\in\mathcal{V}}
 \Gbf_{i,j}\xbf_{i},
\end{equation}
where the transmitting signals $\xbf_k\in\FF_p^q$, and the ``channel''
matrices $\Gbf_{i,j}\in \FF_p^{q\times q}$. All the operations are
done over the finite field $\FF_p$, and the network $\mathcal{G}$,
implies that $\Gbf_{i,j}=\mathbf{0},  i\notin\mathcal{N}_j$ reducing
the sum in (\ref{eq:LinDetModel}) from $N=|\mathcal{V}|$ terms {\em i.e.,}
all transmitting nodes in the network, to just the input neighbors of
$j$.

If we look at the cut-set upper bound for general deterministic
networks (\ref{eq:CutSet}), it is easy to see in a special case of
linear finite-field deterministic networks that all cut values are
simultaneously optimized by independent and uniform distribution of
$\{x_i\}_{i\in\mathcal{V}}$. Moreover the optimum value of each cut
$\Omega$ is logarithm of the size of the range space of the transfer
matrix $\Gbf_{\Omega,\Omega^c}$ associated with that cut, {\em i.e.,}
the matrix relating the super-vector of all the inputs at the nodes in
$\Omega$ to the super-vector of all the outputs in $\Omega^c$ induced
by (\ref{eq:LinDetModel}). This yields the following complete
characterization as the corollaries of theorem \ref{thm:GenDetNet} and
\ref{thm:GenDetNetMulticast}:
\begin{corollary}
\label{thm:LinDetNet}
Given a linear finite-field relay network (with broadcast and multiple
access), the capacity $C$ of such a relay network is given
by,
\begin{eqnarray}
\label{eq:ThmLinDetNet}
C = \min_{\Omega\in\Lambda_D}
\mathrm{rank}(\Gbf_{\Omega,\Omega^c})\log p.
\end{eqnarray}
\end{corollary}

\begin{corollary}
\label{thm:LinDetNetMulticast}
Given a linear finite-field relay network (with broadcast and multiple
access), the multicast capacity $C$ of such a relay network is given
by,
\begin{eqnarray}
\label{eq:ThmLinDetNetMulticast}
C = \min_{D\in\mathcal{D}}\min_{\Omega\in\Lambda_D}
\mathrm{rank}(\Gbf_{\Omega,\Omega^c})\log p.
\end{eqnarray}
\end{corollary}

For a single source-destination pair the result in Corollary
\ref{thm:LinDetNet} generalizes the classical max-flow min-cut theorem
for wireline networks and for multicast, the result in Corollary
\ref{thm:LinDetNetMulticast} generalizes the
network coding result in \cite{ACLY00} where in both these earlier
results, the communication links are orthogonal. Moreover, as we will
see in the proof, the encoding functions at the relay nodes could be
restricted to linear functions to obtain the result in Corollary
\ref{thm:LinDetNet}.

\subsection{Proof Strategy}

Theorem \ref{thm:GenDetNet} is the main result of the paper and the
rest of the paper is devoted to proving it. First we focus on networks
that have a layered structure, i.e. all paths from the source to the
destination have equal lengths. With this special structure we get a
major simplification: a sequence of messages can each be encoded into
a block of symbols and the blocks do not interact with each other as
they pass through the relay nodes in the network. The proof of the
result for layered network is similar in style to the random coding
argument in \cite{ACLY00}. We do this in sections \ref{secPfIdea},
\ref{sec:LayFF} and \ref{sec:GenDet}, first for the linear
finite-field model in \ref{secPfIdea} and \ref{sec:LayFF} and then for
the general deterministic model in \ref{sec:GenDet}.  Second, we
extend the result to an arbitrary network by considering its
time-expanded representation.  The time-expanded network is layered
and we can apply our result in the first step to it. To complete the
proof of the result, we need to establish a connection between the cut
values of the time-expanded network and those of the original
network. We do this using sub-modularity properties of entropy in
Section \ref{sec:TimExp}\footnote{The concept of time-expanded
representation is also used in \cite{ACLY00}, but the use there is to
handle cycles. Our main use is to handle interaction between messages
transmitted at different times, an issue that only arises when there
is interference at nodes.}.

%%To make the proofs more readable the result for layered networks are proved first for linear %%deterministic model that is more intuitive in section \ref{sec:LayFF} and then for  general %%deterministic model in section \ref{sec:GenDet}. Finally we will discuss time-expansion %%ideas and prove the results for arbitrary networks in section \ref{subsec:TimExp}.

\section{Linear Model: An Example}
\label{secPfIdea}
In this section we give the encoding scheme for the linear
deterministic model of (\ref{eq:LinDetModel}) in Section
\ref{subsec:EncLinDet}. In Section \ref{subsec:PfIdea} we illustrate
the proof techniques on a simple linear unicast relay network example.

\subsection{Encoding for linear deterministic model}
\label{subsec:EncLinDet}
%%At a particular time $t$, the deterministic model in
%%(\ref{eq:LinDetModel}) implies that
%%\[
%%\ybf_{j}^{[t]}=\sum_{k\in\mathcal{V}}
%% \Gbf_{i,j}\xbf_{i}^{[t]},\,\, t=1,2,\ldots,T.
%%\]
We have a single source $S$ with message $W\in\{1,2,\ldots,2^{TKR}\}$
which is encoded by the source $S$ into a signal over $KT$
transmission times (symbols), giving an overall transmission rate of
$R$. Each relay operates over blocks of time $T$ symbols, and uses a
mapping $f_j^{(k)}:\mathcal{Y}_j^T\rightarrow \mathcal{X}_j^T$ its
received symbols from the previous block of $T$ symbols to transmit
signals in the next block. In particular, block $k$ of $T$ received
symbols is denoted by
$\ybf_j^{(k)}=\{\ybf_j^{[(k-1)T+1]},\ldots,\ybf_j^{[kT]}\}$ and the
transmit symbols by $\xbf_j^{(k)}$. For the model
(\ref{eq:LinDetModel}), we will use linear mappings $f_j(\cdot)$, {\em
i.e.,}
\begin{equation}
\label{eq:LinEncFn}
\xbf_j^{(k)} = \Fbf_j^{(k)}\ybf_j^{(k-1)},
\end{equation}
where $\Fbf_j^{(k)}$ is chosen uniformly randomly over all matrices in
$\FF_p^{q\times q}$.  Each relay does the encoding prescribed by
(\ref{eq:LinEncFn}). Given the knowledge of all the encoding functions
$\Fbf_j$ at the relays and signals received over $K+|\mathcal{V}|-2$
blocks, the decoder $D\in\mathcal{D}$, attempts to decode the message
$W$ sent by the source.

\subsection{Proof illustration}
\label{subsec:PfIdea}

In order to illustrate the proof ideas of Theorem
(\ref{thm:GenDetNet}) we examine the network shown in Figure
\ref{fig:Confusability}. We will analyze this network first for linear
deterministic model and then we use the same example to illustrate the
ideas for general deterministic functions in Section
\ref{subsec:PfIdeaGen}.

\begin{figure}[h]
\centering
\input{Confusability.pstex_t}
\label{fig:Confusability}
\end{figure}

The network given in Figure \ref{fig:Confusability} is an example of a
{\em layered} network where the number of ``hops'' for each path from
$S$ to $D$ is equal to $3$ in this case\footnote{Note that in the
equal path network we do not have ``self-interference'' since all
path-lengths from $S$ to $D$ in terms of ``hops'' are equal, though as
we will see in the analysis that can easily be taken care of. However
we do allow for self-interference in the model and we choose to handle
such loops, and more generally cyclic networks, through time-expansion
as will be seen in Section \ref{sec:TimExp}.}. The key simplification
that occurs for layered networks is that we can divide the message $W$
into $K$ parts (sub-messages), each taking values in
$w_k\in\{1,2,\ldots,2^{TR}\},k=1,\ldots,K$. By doing this in Figure
\ref{fig:Confusability}, we see that for example, nodes $A_1,A_2$ are
sending signals which pertain to the same sub-message
$w_k$. Therefore, the ``interfering'' signals in node $B_1$ are both
about the same sub-message. This is a statement that holds in general
for layered networks. For example in block number $k=3$, the source is
sending a signal about $w_3$, $A_1,A_2$ are sending signals that
depend on $w_2$ and $B_1,B_2$ in turn are sending a signal to $D$
which depends on $w_1$. This message synchronization implies that we
can focus our attention on the error probability of a single
sub-message $w=w_1$ without loss of generality.

Now, since we have a deterministic network, the message $w$ will be
mistaken for another message $w'$ is if the received signal
$\ybf_D^{(3)}(w)$ under $w$, is the same as that would have been
received under $w'$. This leads to a notion of {\em
distinguishability}, which is that messages $w,w'$ are distinguishable
at any node $j$ if $\ybf_j(w)\neq \ybf_j(w')$.

The probability of error at decoder $D$ can be upper bounded using the
union bound as,
\begin{equation}
\label{eq:PeUB}
P_e \leq 2^{RT}\prob{w\rightarrow w'}
= 2^{RT}\prob{\ybf_D^{(3)}(w)=\ybf_D^{(3)}(w')} .
\end{equation}
For the deterministic network, this event, is random only due to the
randomness in the encoder map. Therefore, the probability of this
event depends on the probability that we choose such an encoder map.
Now, we can write,
{\footnotesize
\begin{eqnarray}
\nonumber
& \displaystyle \prob{w\rightarrow w'} = \sum_{\Omega\in\Lambda_D}
\\ \label{eq:PwErr}& \underbrace{\prob{\mbox{Nodes in } \Omega
\mbox{ can distinguish } w,w' \mbox{ and nodes in } \Omega^c \mbox{
cannot} }}_{\mathcal{P}} ~
\end{eqnarray}
}
since the events that correspond to occurrence of the
distinguishability sets $\Omega\in\Lambda_D$ are disjoint. Let us
examine one term in the summation in (\ref{eq:PwErr}). The
distinguishability of $w=w_1$ from $w'=w_1'$ for the nodes $A_1,A_2$
are from signals $\ybf_{A_1}^{(1)},\ybf_{A_2}^{(1)}$, for the nodes
$B_1,B_2$ are from signals $\ybf_{B_1}^{(2)},\ybf_{B_2}^{(2)}$ and for
the receiver $D$ it is $\ybf_D^{(3)}(w)$. For notational simplicity we
will drop the block numbers associated with the transmitted and
received signals for this analysis.

For the cut $\Omega=\{S,A_1,B_1\}$, a necessary condition for the
distinguishability set to be this cut is that
$\ybf_{A_2}(w)=\ybf_{A_2}(w')$, along with
$\ybf_{B_2}(w)=\ybf_{B_2}(w')$ and $\ybf_{D}(w)=\ybf_{D}(w')$.  Since
the source does a random linear mapping of the message onto
$\xbf_S(w)$, the probability that $\ybf_{A_2}(w)=\ybf_{A_2}(w')$ is
given by, {\small
\begin{equation}
\label{eq:ExLinEP1}
\prob{(\Ibf_T\otimes\Gbf_{S,A_2})(\xbf_S(w)-\xbf_S(w'))=\mathbf{0}} =
p^{-T\mbox{rank}(\Gbf_{S,A_2})},
\end{equation}
} since the random mapping given in (\ref{eq:LinEncFn}) induces
independent uniformly distributed $\xbf_S(w),\xbf_S(w')$. Here,
$\otimes$ is the Kronecker matrix product.  Now, in order to analyze
the probability that $\ybf_{B_2}(w)=\ybf_{B_2}(w')$, we see that since
$\ybf_{A_2}(w)=\ybf_{A_2}(w')$, $\xbf_{A_2}(w)=\xbf_{A_2}(w')$, {\em
i.e.,} the {\em same} signal is sent under both $w,w'$. Therefore, we
get the probability of $\ybf_{B_2}(w)=\ybf_{B_2}(w')$ given that the
distinguishability set is $\Omega=\{S,A_1,B_1\}$, as,
\begin{footnotesize}
\begin{equation}
\label{eq:ExLinEP2}
\prob{(\Ibf_T\otimes\Gbf_{A_1,B_2})(\xbf_{A_1}(w)-\xbf_{A_1}(w'))=\mathbf{0}}
= p^{-T\mbox{rank}(\Gbf_{A_1,B_2})}.
\end{equation}
\end{footnotesize}
Similarly we get,
{\footnotesize
\begin{eqnarray}
\nonumber &&
\prob{\ybf_{D}(w)=\ybf_{D}(w')|\mbox{distinguishability set } \Omega}  \\ \nonumber && \quad \quad \quad =
\prob{(\Ibf_T\otimes\Gbf_{B_1,D})(\xbf_{B_1}(w)-\xbf_{B_1}(w'))=\mathbf{0}}
\\ \label{eq:ExLinEP3} &&  \quad \quad \quad  = p^{-T\mbox{rank}(\Gbf_{B_1,D})}.
\end{eqnarray}
}
Putting these together, since all three would need to occur, we see
that in (\ref{eq:PwErr}), for the network in Figure
\ref{fig:Confusability}, we have,
\begin{eqnarray}
\nonumber \mathcal{P} & \leq &
p^{-T\mbox{rank}(\Gbf_{S,A_2})}p^{-T\mbox{rank}(\Gbf_{A_1,B_2})}
p^{-T\mbox{rank}(\Gbf_{B_1,D})}\\ \label{eq:ConfProb} &=& p^{-T\{\mbox{rank}(\Gbf_{S,A_2})
+\mbox{rank}(\Gbf_{A_1,B_2})+\mbox{rank}(\Gbf_{B_1,D})\}}.
\end{eqnarray}
Note that since in this example,
\[
\Gbf_{\Omega,\Omega^c} = \left [ \begin{array}{ccc}\Gbf_{S,A_2} &
\Zerobf & \Zerobf \\ \Zerobf  & \Gbf_{A_1,B_2} & \Zerobf \\
\Zerobf & \Zerobf & \Gbf_{B_1,D} \end{array}\right ],
\]
the upper bound for $\mathcal{P}$ in (\ref{eq:ConfProb}) is
exactly $2^{-T\mbox{rank}(\Gbf_{\Omega,\Omega^c})}$. Therefore,
by substituting this back into (\ref{eq:PwErr}) and (\ref{eq:PeUB}),
we see that
\begin{equation}
\label{eq:RateBndLinEx}
\displaystyle
P_e \leq 2^{RT} |\Lambda_D| p^{-T\min_{\Omega\in\Lambda_D}
\mbox{rank}(\Gbf_{\Omega,\Omega^c})},
\end{equation}
which can be made as small as desired if $R<\min_{\Omega\in\Lambda_D}
\mbox{rank}(\Gbf_{\Omega,\Omega^c})\log p$, which is the result claimed in
Corollory \ref{thm:LinDetNet}.

These ideas motivate first focussing on layered networks
as done in Section \ref{sec:LayFF}. The major simplification that we
get in this case is that the signals associated with different
messages do not get mixed in the network and hence we can only focus
on one message. Note that another simplification in layered (equal
path) networks is that for a given node $j$, it is enough to choose
the same encoding function $f_j$ for each block $k$.

Now the general result for layered networks are proved in two parts: first for linear deterministic model and then for general deterministic model.

\section{Layered networks: linear deterministic model}
\label{sec:LayFF}
In this section we prove main corollaries \ref{thm:LinDetNet} and
\ref{thm:LinDetNetMulticast} for layered networks.  In a layered
network, for each node $j$ we have a length $l_j$ from the source and
all the incoming signals to node $j$ are from nodes $i$ whose distance
from the source are $l_i=l_j-1$. Therefore, as in the example network
of Figure \ref{fig:Confusability}, we see that there is message
synchronization, {\em i.e.,} all signals arriving at node $j$ are
encoding the same sub-message.

Suppose message $w_k$ is sent by the source in block $k$, then since
each relay $j$ operates only on block of lengths $T$, the signals
received at block $k$ at any relay pertain to only message $w_{k-l_j}$
where $l_j$ is the path length from source to relay $j$.  To
explicitly indicate this we denote by $\ybf_j^{(k)}(w_{k-l_j})\in
\FF_{p}^{qT}$ as the received signal at block $k$ at node $j$. We
also denote the transmitted signal at block $k$ as
$\xbf_j^{(k)}(w_{k-1-l_j})\in\FF_{p}^{qT}$ which is obtained by
randomly mapping $\ybf_j^{(k-1)}(w_{k-1-l_j})\in\FF_{p}^{qT}$.

Since we have a layered network, without loss of generality
consider the message $w=w_1$ transmitted by the source at block
$k=1$. At node $j$ the signals pertaining to this message are received
by the relays at block $l_j$. We analyze a $l_D$-layer network, each
layer is a MIMO sub-network.  Therefore, as in the analysis of
(\ref{eq:PwErr}), we see that
{\footnotesize
\begin{eqnarray}
\nonumber
& P_e^{(D)} \leq 2^{RT}\sum_{\Omega\in\Lambda_D}\\ \label{eq:LayUB} & \underbrace{\prob{\mbox{Nodes
in } \Omega \mbox{ can distinguish } w,w' \mbox{ and nodes in }
\Omega^c \mbox{ cannot} }}_{\mathcal{P}} ~~
\end{eqnarray}
}

We define $\Gbf_{\Omega,\Omega^c}$ as the transfer matrix associated with the
nodes in $\Omega$ to the nodes in $\Omega^c$. Note that
since we have a layered network this transfer matrix breaks up
into block diagonal elements corresponding to each of the $l_D$ layers
of the network. More precisely, we can create $d=l_D$ disjoint
sub-networks of nodes corresponding to each layer of the network, with
$\beta_l(\Omega)$ nodes at distance $l-1$ from $S$ that are in $\Omega$, on one side and
$\gamma_l(\Omega)$ nodes at distance $l$ from $S$ that are in $\Omega^c$, on the other, for
$l=1,\ldots,l_D$.

Each node $i\in\beta_l(\Omega)$ sees a signal related to $w=w_1$ in
block $l_i=l-1$, and therefore waits to receive this block and then
does a random mapping to $\xbf_i^{(l_i)}(w)\in\FF_{p}^{qT}$ The random
mapping is done as in (\ref{eq:LinEncFn}), by choosing a random matrix
$\Fbf_i$ of size $Tq\times Tq $ and creating
\begin{eqnarray}
\label{eq:RandomBinning}
\xbf_i^{(l_i)}(w) = \Fbf_i \ybf_i^{(l_i-1)}(w)
\end{eqnarray}

The received signals in the nodes $j\in\gamma_l(\Omega)$ are linear
transformations of the transmitted signals from nodes
$\mathcal{T}_l=\{u: (u,v)\in \mathcal{E}, v\in\gamma_l(\Omega)\}$. That
is, its output depends not only on the transmitters in $\beta_l$, but
also other transmitters at distance $l-1$ from $S$ that are part of
$\Omega^c$.  Since all the receivers in $\gamma_l$ are at distance $l$
from $S$, they form the receivers of the MIMO layer $l$, and we denote
this vector received signal as $\zbf_l(w)$, and this can be done for
all layers $l=1,\ldots,l_D$.  Note that as in the example network of
Section \ref{subsec:PfIdea}, for all the transmitting nodes in
$\mathcal{T}$ which cannot distinguish between $w,w'$ the transmitted
signal would be the same under both $w$ and $w'$.  Therefore, in order
to calculate the probability that nodes in $\gamma_l$ cannot
distinguish between $w,w'$ or that $\zbf_l(w)-\zbf_l(w')=\mathbf{0}$,
we see that
\begin{equation}
\zbf_l(w)-\zbf_l(w') = \tilde{\Gbf}_l\left [\ubf_l(w)-\ubf_l(w')\right
], \,\, l=1,\ldots,d
\end{equation}
where the transmitted signals from $\beta_1,\ldots,\beta_d$ are
clubbed together\footnote{Just as in the received signals, in clubbing
together the transmitted signals into $\ubf_l(w)$, we put together
signals transmitted at the \emph{same} time instant together. This can
be done since we have broken the network into the clusters/stages with
identical path lengths.} and denoted by $\ubf_l(w),l=1,\ldots,d$.
Also, due to the time-invariant channel conditions we see that
$\tilde{\Gbf}_l=\Ibf_T\otimes\Gbf_l$, where $\otimes$ is the Kronecker
product.  Since we are trying to calculate the probability that
$\zbf_l(w)=\zbf_l(w'), l=1,\ldots,d$, and hence we need to find the
probability that $\ubf_l(w)-\ubf_l(w')$ lies in the null space of
$\Gbf_l$ for {\em each} $l=1,\ldots,d$.

Now, if the distinct signals $\ybf_i^{(l_i)}(w),\ybf_i^{(l_i)}(w')$
received at the nodes $i\in\beta_l$ could be {\em jointly uniformly
and independently} mapped to the transmitted signals
$\ubf_l(w),\ubf_l(w')$, then we could say that the probability of this
occurrence is $\frac{\mbox{size of null space}}{\mbox{size of whole
space}}$. Clearly this is given by,
{\small
\begin{equation}
\label{eq:PEP}
\prob{\ubf_l(w)-\ubf_l(w')\in\mathcal{N}(\tilde{\Gbf}_l)}=p^{-\mathrm{rank}(\tilde{\Gbf}_l)}
=p^{-T\mathrm{rank}(\Gbf_l)}.
\end{equation}
}
However, even though the signals $\ybf_i^{(l_i)}(w)$ are uniformly
randomly mapped {\em individually} at each node $i\in\beta_l$, the
overall map across all nodes in $\beta_l$ is also uniform, and hence
the probability given in (\ref{eq:PEP}) is the correct one. Since the
events in each of the stages/clusters are independent, we get that
{\footnotesize
\begin{align*}
&\prob{\ubf_l(w)-\ubf_l(w')\in\mathcal{N}(\tilde{\Gbf}_l),
l=1,\ldots,d} = \prod_{l=1}^dp^{-\mathrm{rank}(\tilde{\Gbf}_l)} \\
&= p^{-T\sum_{l=1}^d\mathrm{rank}(\Gbf_l)} ~~~~~
\label{eq:PEP1}
\end{align*}
}

Therefore, we see that
\begin{equation}
\label{eq:PEP2}
\mathcal{P} %=\prob{\mathcal{S}(w,w')=\Omega}
\leq p^{-T\sum_{l=1}^d\mathrm{rank}(\Gbf_l)}.
\end{equation}
Now the probability of mistaking $w$ for $w'$ at receiver  $D\in\mathcal{D}$ is therefore
\begin{eqnarray}
\nonumber
\displaystyle
\prob{w\rightarrow w'}
&\leq &
\sum_{\Omega\in\Lambda_D}
p^{-T\sum_{l=1}^{d(\Omega)}\mathrm{rank}(\Gbf_l(\Omega))}
\\ \nonumber
&\leq& 2^{|\mathcal{V}|}
p^{-T\min_{\Omega\in\Lambda}\mathrm{rank}(\Gbf_{\Omega,\Omega^c})},
\end{eqnarray}
where we have used $|\Lambda_D|\leq 2^{|\mathcal{V}|}$.  Note that we
have used the fact that since $\Gbf_{\Omega,\Omega^c}$ was block diagonal, with
blocks, $\Gbf_l(\Omega)$, we see that
$\sum_{l=1}^{d(\Omega)}\mathrm{rank}(\Gbf_l(\Omega))=\mathrm{rank}(\Gbf_{\Omega,\Omega^c})$.
If we declare an error if {\em any} receiver $D\in\mathcal{D}$ makes
an error, we see that since we have $2^{RT}$ messages, from the union
bound we can drive the error probability to zero if we have,
\begin{equation}
\label{eq:LayRateBnd}
\displaystyle R <
\min_{D\in\mathcal{D}}\min_{\Omega\in\Lambda_D}\mathrm{rank}(\Gbf_{\Omega,\Omega^c})
\log p.
\end{equation}
Therefore for the layered (equal path) network with linear
deterministic functions, since as seen in Section \ref{sec:MainRes},
the cut-set is also identical to the expression in
(\ref{eq:LayRateBnd}), we have proved the following result.
\begin{theorem}
\label{thm:StagLinDetNet}
Given a layered (equal path) linear finite-field relay network (with
broadcast and multiple access), the multicast capacity $C$ of such a
relay network is given by,
\begin{eqnarray}
\label{eq:ThmLinDetStagNet}
C = \min_{D\in\mathcal{D}} \min_{\Omega\in\Lambda_D}
\mathrm{rank}(\Gbf_{\Omega,\Omega^c})\log p,
\end{eqnarray}
\end{theorem}

\section{Layered networks: general deterministic model}
\label{sec:GenDet}
In this section we prove main theorems \ref{thm:GenDetNet} and \ref{thm:GenDetNetMulticast} for layered networks. We first generalize the encoding scheme to accommodate arbitrary deterministic
functions of (\ref{eq:GenDetModel}) in Section \ref{subsec:EncGenDet}.
We then illustrate the ingredients of the proof using the same example
as in Section \ref{subsec:PfIdea}.  Then we prove the result for
layered networks in Section \ref{subsec:LayGen}.

\subsection{Encoding for general  deterministic model}
\label{subsec:EncGenDet}
We assume a clocked network as in Section
\ref{subsec:EncLinDet}. Therefore, for such a clocked network, the
deterministic model in (\ref{eq:LinDetModel}) implies that
\[
\ybf_{j}^{[t]}=g_j(\{x_i^{[t]}\}_{i\in\mathcal{N}_j}),\,\, t=1,2,\ldots,T.
\]
We have a single source $S$ with message $W\in\{1,2,\ldots,2^{TKR}\}$
which is encoded by the source $S$ into a signal over $KT$
transmission times (symbols), giving an overall transmission rate of
$R$. We will use strong (robust) typicality as defined in \cite{OR01}.
The notion of joint typicality is naturally extended from Definition
\ref{def:RobTyp}.
\begin{defn}
\label{def:RobTyp}
We define $\underline{x}\in T_{\delta}$ if
\[
|\nu_{\underline{x}}(x)-p(x)| \leq \delta p(x),
\]
where $\nu_{\underline{x}}(x)=\frac{1}{T}|\{t: x_t=x\}|$, is the
empirical frequency.
\end{defn}

Each relay operates over blocks of time $T$ symbols, and uses a
mapping $f_j^{[t]}:\mathcal{Y}_j^T\rightarrow \mathcal{X}_j^T$ its
received symbols from the previous block of $T$ symbols to transmit
signals in the next block. In particular, block $k$ of $T$ received
symbols is denoted by
$\ybf_j^{(k)}=\{y^{[(k-1)T+1]},\ldots,y^{[kT]}\}$ and the transmit
symbols by $\xbf_j^{(k)}$.  Choose some product distribution
$\prod_{i\in\mathcal{V}}p(x_i)$. At the source $S$, map each of the
indices in $W\in\{1,2,\ldots,2^{TKR}\}$ choose $f_S^{(k)}(W)$ onto a
sequence uniformly drawn from $T_{\delta}(X_S)$, which is the typical
set of sequences in $\mathcal{X}_S^T$.  At any relay node $j$ choose
$f_j^{(k)}$ to map each typical sequence in $\mathcal{Y}_j^T$ {\em
i.e.,} $T_{\delta}(Y_j)$ onto typical set of transmit sequences {\em
i.e.,} $T_{\delta}(X_j)$, as
\begin{equation}
\label{eq:GenEncFn}
\xbf_j^{(k)} = f_j^{(k)}(\ybf_j^{(k-1)}),
\end{equation}
where $f_j^{(k)}$ is chosen to map uniformly randomly each sequence in
$T_{\delta}(Y_j)$ onto $T_{\delta}(X_j)$ and is done independently for
each block $k$.  Each relay does the encoding prescribed by
(\ref{eq:GenEncFn}). Given the knowledge of all the encoding functions
$f_j^{(k)}$ at the relays and signals received over
$K+|\mathcal{V}|-2$ blocks, the decoder $D\in\mathcal{D}$, attempts to
decode the message $W$ sent by the source.

\subsection{Proof illustration}
\label{subsec:PfIdeaGen}

Now, we illustrate the ideas behind the proof of Theorem
\ref{thm:GenDetNet} for layered networks using the same example as in
Section \ref{subsec:PfIdea}, which was done for the linear
deterministic model.  Since we are dealing with deterministic
networks, the logic upto (\ref{eq:PwErr}) in Section
\ref{subsec:PfIdea} remains the same. We will again illustrate the
ideas using the cut $\Omega=\{S,A_1,B_1\}$. As in Section
\ref{subsec:PfIdea}, necesary condition for this set to be the
distinguishability set is that $\ybf_{A_2}(w)=\ybf_{A_2}(w')$, along
with $\ybf_{B_2}(w)=\ybf_{B_2}(w')$ and $\ybf_{D}(w)=\ybf_{D}(w')$.
Notice that as in Section \ref{subsec:PfIdea}, we are suppressing the
block numbers associated with the received signals. It is clear that
for $w=w_1$, the block numbers associated with
$\ybf_{A_2},\ybf_{B_2},\ybf_{D}$ are $1,2,3$ respectively.

Note that since $\ybf_j\in T_{\delta}(Y_j)$ with high probability, we
can focus only on the typical received signals. Let us first examine
the probability that $\ybf_{A_2}(w)=\ybf_{A_2}(w')$. Since $S$ can
distinguish between $w,w'$, it maps these sub-messages independently
to two transmit signals $\xbf_S(w),\xbf_S(w')\in T_{\delta}(X_S)$,
hence we can see that this probability is,
\begin{equation}
\label{eq:ExGenEP1}
\prob{(\xbf_S(w'),\ybf_{A_2}(w))\in T_{\delta}(X_S,Y_{A_2})} =
2^{-TI(X_S;Y_{A_2})}.
\end{equation}
 Now, in order to analyze the probability that
$\ybf_{B_2}(w)=\ybf_{B_2}(w')$, as seen in the linear model analysis,
we see that since $\ybf_{A_2}(w)=\ybf_{A_2}(w')$,
$\xbf_{A_2}(w)=\xbf_{A_2}(w')$, {\em i.e.,} the {\em same} signal is
sent under both $w,w'$. Therefore, since naturally
$(\xbf_{A_2}(w),\ybf_{B_2}(w))\in T_{\delta}(X_{A_2},Y_{B_2})$,
obviously, $(\xbf_{A_2}(w'),\ybf_{B_2}(w))\in
T_{\delta}(X_{A_2},Y_{B_2})$ as well.  Therefore, under $w'$, we
already have $\xbf_{A_2}(w')$ to be jointly typical with the signal
that is received under $w$. However, since $A_1$ can distinguish
between $w,w'$, it will map the transmit sequence $\xbf_{A_1}(w')$ to
a sequence which is independent of $\xbf_{A_1}(w)$ transmitted under
$w$. Since an error occurs when
$(\xbf_{A_1}(w'),\xbf_{A_2}(w'),\ybf_{B_2}(w))\in
T_{\delta}(X_{A_1},X_{A_2},Y_{B_2})$, and since $A_2$ cannot
distinguish between $w,w'$, we also have
$\xbf_{A_2}(w)=\xbf_{A_2}(w')$, we require that
$(\xbf_{A_1},\xbf_{A_2},\ybf_{B_2})$ generated like
$p(\xbf_{A_1})p(\xbf_{A_2},\ybf_{B_2})$ behaves like a jointly typical
sequence. Therefore, this probability is given by,
\begin{small}
\begin{eqnarray}
\nonumber & \prob{(\xbf_{A_1}(w'),\xbf_{A_2}(w),\ybf_{B_2}(w))\in
T_{\delta}(X_{A_1},X_{A_2}Y_{B_2})} \stackrel{\cdot}{=} \\ \label{eq:ExGenEP2} &
2^{-TI(X_{A_1};Y_{B_2},X_{A_2})} \stackrel{(a)}{=}
2^{-TI(X_{A_1};Y_{B_2}|X_{A_2})},
\end{eqnarray}
\end{small}
where $\stackrel{\cdot}{=}$ indicates exponential equality (where we
neglect subexponential constants), and $(a)$ follows since we have
generated the mappings $f_j$ independently, it induces an independent
distribution on $X_{A_1},X_{A_2}$. Another way to see this is that the
probability of (\ref{eq:ExGenEP2}) is given by
$\frac{|T_{\delta}(\Xbf_{A_1}|\xbf_{A_2},\ybf_{B_2})|}{|T_{\delta}(\Xbf_{A_1})|}$,
which by using properties of (robustly) typical sequences \cite{OR01}
yields the same expression as in (\ref{eq:ExGenEP2}). Note that the
calculation in (\ref{eq:ExGenEP2}) is similar to one of the error
event calculations in a multiple access channel,

Using a similar logic we can write,
\begin{eqnarray}
\nonumber &
\prob{(\xbf_{B_1}(w'),\xbf_{B_2}(w),\ybf_{D}(w))\in
T_{\delta}(X_{B_1},X_{B_2}Y_{D})} \stackrel{\cdot}{=}
\\ \label{eq:ExGenEP3} & 2^{-TI(X_{B_1};Y_{D},X_{B_2})} \stackrel{(a)}{=}
2^{-TI(X_{B_1};Y_{D}|X_{B_2})}.
\end{eqnarray}
Therefore, putting (\ref{eq:ExGenEP1})--(\ref{eq:ExGenEP3}) together as
done in (\ref{eq:ConfProb}) we get
\[
\mathcal{P}\leq
2^{-T\{I(X_S;Y_{A_2})+I(X_{A_1};Y_{B_2}|X_{A_2})+I(X_{B_1};Y_{D}|X_{B_2})\}}
\]
Note that for this example, due to the Markovian structure of the
network we can see that\footnote{Note that though in the encoding
scheme there is a dependence between $X_{A_1},X_{A_2},X_{B_1},X_{B_2}$
and $X_S$, in the single-letter form of the mutual information, under
a product distribution, $X_{A_1},X_{A_2},X_{B_1},X_{B_2},X_S$ are
independent of each other. Therefore for example, $Y_{B_2}$ is
independent of $X_{B_2}$ leading to
$H(Y_{B_2}|X_{A_2},X_{B_2})=H(Y_{B_2}|X_{A_2})$.  Using this argument for
the cut-set expression $I(Y_{\Omega^c};X_{\Omega}|X_{\Omega^c})$, we get
the expansion.}  $I(Y_{\Omega^c};X_{\Omega}|X_{\Omega^c})
=I(X_S;Y_{A_2})+I(X_{A_1};Y_{B_2}|X_{A_2})+I(X_{B_1};Y_{D}|X_{B_2})$,
hence as in (\ref{eq:RateBndLinEx}) we get that,
\begin{equation}
\label{eq:RateBndGenEx}
\displaystyle
P_e \leq 2^{RT} |\Lambda_D| 2^{-T\min_{\Omega\in\Lambda_D}
I(Y_{\Omega^c};X_{\Omega}|X_{\Omega^c})},
\end{equation}
and hence the error probability can be made as small as desired if
$R<\min_{\Omega\in\Lambda_D}H(Y_{\Omega^c}|X_{\Omega^c})$, since we
are dealing with deterministic networks.

\subsection{General deterministic model: Proof for layered networks}
\label{subsec:LayGen}

As in the example illustrating the proof in Section
\ref{subsec:PfIdeaGen}, the logic of the proof in the general
deterministic functions follows that of the linear model quite
closely. In particular, as in Section \ref{sec:LayFF} we can define
the bi-partite network associated with a cut $\Omega$. Instead of a
transfer matrix $\Gbf_{\Omega,\Omega^c}(\cdot)$ associated with the cut, we
have a transfer function $\tilde{\Gbf}_{\Omega}$.  Since we are still
dealing with a layered network, as in the linear model case, this
transfer function breaks up into components corresponding to each of
the $l_D$ layers of the network. More precisely, we can create $d=l_D$
disjoint sub-networks of nodes corresponding to each layer of the
network, with $\beta_l(\Omega)$ nodes at distance $l-1$ from $S$, on
one side and $\gamma_l(\Omega)$ nodes at distance $l$ from $S$, on the
other, for $l=1,\ldots,l_D$. Each of this MIMO clusters have a
transfer function $\Gbf_l(\cdot),l=1,\ldots,l_D$ associated with them.

As in the linear model, each node $i\in\beta_l(\Omega)$ sees a signal
related to $w=w_1$ in block $l_i=l-1$, and therefore waits to receive
this block and then does a mapping using the general encoding function
given in (\ref{eq:GenEncFn}) as
\begin{equation}
\label{eq:GenEncFnRep}
\xbf_j^{(k)}(w) = f_j^{(k)}(\ybf_j^{(k-1)}(w)).
\end{equation}
The received signals in the nodes $j\in\gamma_l(\Omega)$ are
deterministic transformations of the transmitted signals from nodes
$\mathcal{T}_l=\{u: (u,v)\in \mathcal{E}, v\in\gamma_l(\Omega)\}$. As
in the linear model analysis of Section \ref{sec:LayFF}, the
dependence is on all the transmitting signals at distance $l-1$ from
the source, not just the ones in $\beta_l\subset\Omega$.  Since all
the receivers in $\gamma_l$ are at distance $l$ from $S$, they form
the receivers of the MIMO layer $l$, and we denote this vector
received signal as $\zbf_l(w)$, and this can be done for all layers
$l=1,\ldots,l_D$.  Note that as in the example network of Section
\ref{subsec:PfIdeaGen}, for all the transmitting nodes in
$\mathcal{T}$ which cannot distinguish between $w,w'$ the transmitted
signal would be the same under both $w$ and $w'$.  Therefore, all the
nodes in $\mathcal{T}_l\cap \Omega^c$ cannot distinguish between
$w,w'$ and therefore
\[
\xbf_j(w)=\xbf_j(w'),\,\,\,j\in \mathcal{T}_l\cap \Omega^c.
\]
Hence it is clear that since $(\{\xbf_j(w)\}_{j\in
\mathcal{T}_l\cap \Omega^c},\zbf_l(w))\in T_{\delta}$, we have that
\[
(\{\xbf_j(w')\}_{j \in \mathcal{T}_l\cap \Omega^c},\zbf_l(w))\in
T_{\delta}.
\]
Therefore, just as in Section \ref{subsec:PfIdeaGen}, we see that the probability
that $\zbf_l(w)=\zbf_l(w')$, is given by,
\begin{equation}
\label{eq:ClusterProb}
\prob{\zbf_l(w)=\zbf_l(w')} \stackrel{\cdot}{=}
2^{-TI(X_{\mathcal{T}_l\cap \Omega};Z_l,X_{\mathcal{T}_l\cap
\Omega^c})}.
\end{equation}
Since the events in each of the MIMO stages (clusters) are independent, we
get that
\begin{eqnarray}
\nonumber
& \prob{\zbf_l(w)=\zbf_l(w'),l=1,\ldots,d} = \\ \label{eq:GenPEP1} & \prod_{l=1}^d
2^{-TI(X_{\mathcal{T}_l\cap \Omega};Z_l,X_{\mathcal{T}_l\cap
\Omega^c})}  = 2^{-T\sum_{l=1}^d H(Z_l|X_{\mathcal{T}_l\cap
\Omega^c})}. ~~
\end{eqnarray}
Note that due to the Markovian nature of the layered network, we see
that $\sum_{l=1}^d H(Z_l|X_{\mathcal{T}_l\cap
\Omega^c})=H(Y_{\Omega^c}|X_{\Omega^c})$.  From this point onwards the
proof closely follows the steps as in the linear model from
(\ref{eq:PEP2}) onwards.  Therefore for the layered (equal path)
network with general deterministic functions we have proved the
following result. Similarly in multicast scenario  we declare an error if {\em any} receiver $D\in\mathcal{D}$ makes
an error, we see that since we have $2^{RT}$ messages, from the union
bound we can drive the error probability to zero if we have,
\begin{equation}
\label{eq:LayRateBndGen}
\displaystyle
R < \max_{\prod_{i\in\mathcal{V}} p(x_i)} \min_{D\in\mathcal{D}} \min_{\Omega\in\Lambda_D}
H(Y_{\Omega^c}|X_{\Omega^c}).
\end{equation}
Therefore we have proved the following result.
\begin{theorem}
\label{thm:StagGenDetNet}
Given a layered (equal path) general deterministic relay network (with broadcast and
multiple access),  we can achieve any rate $R$ from $S$ multicasting to all destinations $D\in\mathcal{D}$, with $R$ satisfying:
\begin{eqnarray}
\label{eq:GenDetStagNet}
R <  \max_{\prod_{i\in\mathcal{V}} p(x_i)} \min_{D\in\mathcal{D}} \min_{\Omega\in\Lambda_D}
H(Y_{\Omega^c}|X_{\Omega^c})
\end{eqnarray}
\end{theorem}

\section{Arbitrary networks}
\label{sec:TimExp}

\begin{figure*}%[htp]
     \centering \subfigure[An example of general deterministic network ]{
       \input{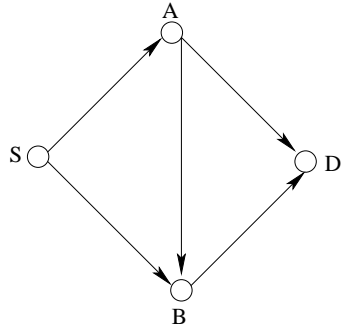}
}
     \hspace{.3in}
     \subfigure[Unfolded deterministic network.
An example of steady cuts and wiggling cuts are respectively shown by solid and dotted lines.]{
       \input{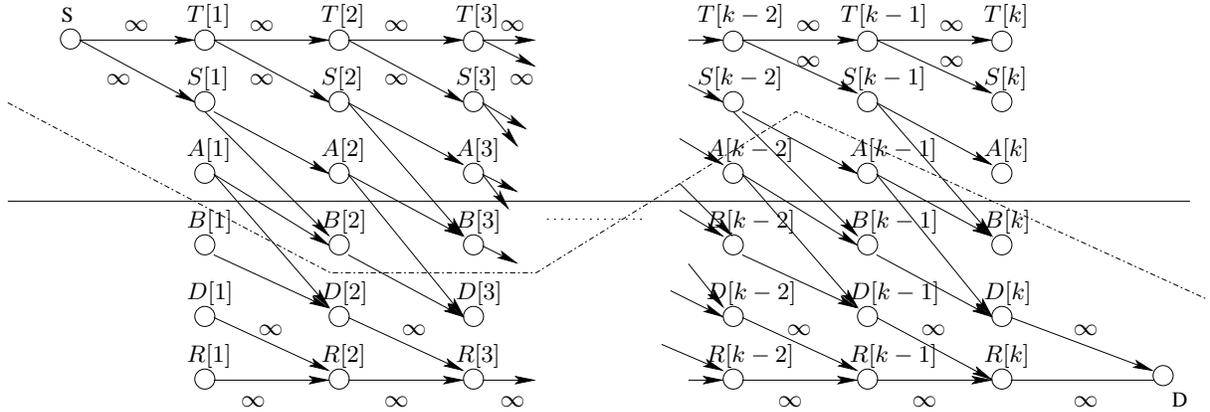}
}
\caption{An example of a general deterministic network with un equal
paths from S to D is shown in $(a)$.  The corresponding unfolded
network is shown in $(b)$.}
\label{fig:unfolding}
\end{figure*}
Given the proof for layered networks with equal path lengths, we are
ready to tackle the proof of Theorem \ref{thm:GenDetNet} and Theorem
\ref{thm:GenDetNetMulticast} for general relay networks.

The ingredients are developed below. First is that any network can be
unfolded over time to create a layered deterministic network (this
idea was introduced for graphs in \cite{ACLY00} to handle cycles in a
graph). The idea is to unfold the network to $K$ stages such that i-th
stage is representing what happens in the network during $(i-1)T$ to
$iT-1$ symbol times. For example in figure \ref{fig:unfolding}(a) a
network with unequal paths from $S$ to $D$ is shown. Figure
\ref{fig:unfolding}(b) shows the unfolded form of this network. As we
notice each node $v \in \mathcal{V}$ is appearing at stage $1 \leq i
\leq K$ as $v[i]$. There are additional nodes: $T[i]$'s and
$R[i]$'s. These nodes are just virtual transmitters and receivers that
are put to buffer and synchronize the network. Since all communication
links connected to these nodes ($T[i]$'s and $R[i]$'s) are modelled as
wireline links without any capacity limit they would not impose any
constraint on the network. One should notice that in general there
must be an infinite capacity link between the same node and itself
appearing at different times however, here we are omitting these links
which means we limit the nodes to have a finite memory $T$. Now we
show the following lemma, \iffalse
\begin{lemma}
\label{lem:Unf}
Assume $\mathcal{G}$ is a linear deterministic network and
$\mathcal{G}_{\text{unf}}^{(K)}$ is a network obtained by unfolding
$\mathcal{G}$ over $K$ time steps (as shown in figure
\ref{fig:unfolding}). Then any communication rate $R$ is
achievable in $\mathcal{G}$ as long as:
\begin{eqnarray}
\label{eq:GeneralAchiRate}
R < \frac{1}{K} \min_{\Omega_{\text{unf}}\in \Lambda_{D,\text{unf}}}
\mathrm{rank}(\Gbf_{\Omega_{\text{unf}},\Omega_{\text{unf}}^c})\log p
\end{eqnarray}
where the minimum is taken over all cuts $\Omega_{\text{unf}}$ in
$\mathcal{G}_{\text{unf}}^{(K)}$.
\end{lemma}
\fi

\begin{lemma}
\label{lem:UnfGen}
Assume $\mathcal{G}$ is a general deterministic network and
$\mathcal{G}_{\text{unf}}^{(K)}$ is a network obtained by unfolding
$\mathcal{G}$ over $K$ time steps (as shown in figure
\ref{fig:unfolding}). Then the following communication rate is
achievable in $\mathcal{G}$:
\begin{eqnarray}
\label{eq:GenNetAchiRate}
R < \frac{1}{K} \max_{\prod_{i\in\mathcal{V}} p(x_i)} \min_{\Omega_{\text{unf}}\in\Lambda_D}
H(Y_{\Omega_{\text{unf}}^c}|X_{\Omega_{\text{unf}}^c})
\end{eqnarray}
where the minimum is taken over all cuts $\Omega_{\text{unf}}$ in
$\mathcal{G}_{\text{unf}}^{(K)}$.
\end{lemma}
\begin{proof}
By unfolding $\mathcal{G}$ we get an acyclic deterministic
network such that all the paths from the source to the destination
have equal length. Therefore by theorem \ref{thm:StagGenDetNet} we can achieve the rate
\beq \displaystyle R_{\text{unf}} <
\max_{\prod_{i\in\mathcal{V}} p(x_i)} \min_{\Omega_{\text{unf}}\in\Lambda_D}
H(Y_{\Omega^c}|X_{\Omega^c})
\eeq
in the time-expanded graph.  Since it takes $K$ steps to translate and achievable scheme in the time-expanded
graph to an achievable scheme in the original graph, then the Lemma is
proved.
\end{proof}

If we look at different cuts in the time-expanded graph we notice that
there are two types of cuts.  One type separates the nodes at
different stages identically. An example of such a steady cut is drawn
with solid line in figure \ref{fig:unfolding} (b) which separates
$\{S,A\}$ from $\{B,D\}$ at all stages. Clearly each steady cut in the
time-expanded graph corresponds to a cut in the original graph and
moreover its value is $K$ times the value of the corresponding cut in
the original network. However there is another type of cut which does
not behave identically at different stages. An example of such a
wiggling cut is drawn with dotted line in figure \ref{fig:unfolding}
(b). There is no correspondence between these cuts and the cuts in the
original network.

Now comparing Lemma \ref{lem:UnfGen} to the main theorem
\ref{thm:GenDetNet} we want to prove, we notice that in this Lemma the
achievable rate is found by taking the minimum of cut-values over all
cuts in the time-expanded graph (steady and wiggling ones). However in
theorem \ref{thm:GenDetNet} we want to prove that we can achieve a
rate by taking the minimum of cut-values over only the cuts in the
original graph or similarly over the steady cuts in the time-expanded
network. So a natural question is that in a time-expanded network does
it make any difference if we take the minimum of cut-values over only
steady cuts rather than all cuts ? Quite interestingly we show in the
following Lemma that asymptotically as $K \rightarrow \infty$ this
difference (normalized by $1/K$) vanishes.

\begin{lemma}
\label{lem:trellis_min_cut} Consider a general deterministic network, $\mathcal{G}$.
Assume a product distribution on $\{x_i\}_{i \in \mathcal{V}}$,
$p(\{x_i\}_{i \in \mathcal{V}})=\prod_{i \in \mathcal{V}} p(x_i)$.
Now in the time-expanded graph, $\mathcal{G}_{\text{unf}}^{(K)}$,
assume that for each node $i \in \mathcal{V}$, $\{x_i[t]\}_{1 \leq t
\leq K}$ are distributed i.i.d. according to $p(x_i)$ in the original
network. Also for any $1 \leq t_1,t_2 \leq K$ and $i \neq j$,
$x_i[t_1]$ is independent of $x_j[t_2]$. Then for any cut
$\Omega_{\text{unf}}$ on the unfolded graph we have, \beq
(K-L+1) \min_{\Omega \in \Lambda_{D}}
H(Y_{\Omega^c}|X_{\Omega^c}) \leq
H(Y_{\Omega_{\text{unf}}^c}|X_{\Omega_{\text{unf}}^c})  \eeq
where $L=2^{|\mathcal{V}|-2}$.
\end{lemma}
Now since for any distribution
\beq
\min_{\Omega_{\text{unf}}\in\Lambda_D}
H(Y_{\Omega_{\text{unf}}^c}|X_{\Omega_{\text{unf}}^c}) \leq K
\min_{\Omega\in\Lambda_D} H(Y_{\Omega^c}|X_{\Omega^c})
\eeq
we have an immediate corollary of this lemma
\begin{corollary}
\label{cor:trellis_min_cut}
Assume $\mathcal{G}$ is a general deterministic network and
$\mathcal{G}_{\text{unf}}^{(K)}$ is a network obtained by unfolding
$\mathcal{G}$ over $K$ time steps then
\begin{eqnarray}
 \nonumber & \lim_{K \rightarrow \infty} \frac{1}{K}
\max_{\prod_{i\in\mathcal{V}} p(x_i)}
\min_{\Omega_{\text{unf}}\in\Lambda_D}
H(Y_{\Omega_{\text{unf}}^c}|X_{\Omega_{\text{unf}}^c}) \\ & =
\max_{\prod_{i\in\mathcal{V}} p(x_i)} \min_{\Omega\in\Lambda_D}
H(Y_{\Omega^c}|X_{\Omega^c}) \end{eqnarray}
\end{corollary}

\vspace{0.1in}
Now by Lemma \ref{lem:UnfGen} and corollary \ref{cor:trellis_min_cut}, the proof of main theorem \ref{thm:GenDetNet} is complete. So we just need to prove Lemma \ref{lem:trellis_min_cut}. First note
that any cut in the unfolded graph, $\Omega_{\text{unf}}$, partitions
the nodes at each stage $1\leq i \leq K$ to $\mathcal{U}_i$ (on the
left of the cut) and $\mathcal{V}_i$ (on the right of the cut). If at
one stage $S[i] \in \mathcal{V}_i$ or $D[i] \in \mathcal{U}_i$ then
the cut passes through one of the infinite capacity edges (capacity
$Kq$) and hence Lemma \ref{lem:trellis_min_cut} is obviously
proved. Therefore without loss of generality assume that $S[i] \in
\mathcal{U}_i$ and $D[i] \in \mathcal{V}_i$ for all $1 \leq i \leq
K$. Now since for each $i \in \mathcal{V}$, $\{x_i[t]\}_{1 \leq t \leq
K}$ are i.i.d distributed we can write\footnote{As in Section \ref{subsec:PfIdeaGen},
under the product distribution the mutual information expression of
the cut-set breaks into a summation.}
\beq
H(Y_{\Omega_{\text{unf}}^c}|X_{\Omega_{\text{unf}}^c})=
\sum_{i=1}^{K-1} H(Y_{\mathcal{V}_{i+1}}|X_{\mathcal{V}_i})
\eeq
For
simplification we define
\beq \psi(\mathcal{V}_1,\mathcal{V}_2)
\triangleq H(Y_{\mathcal{V}_2}|X_{\mathcal{V}_1})
\eeq
then we have
the following lemma, whose proof is in the appendix.
\begin{lemma} \label{lem:loop}
Let $\mathcal{V}_1,\ldots,\mathcal{V}_l$ be $l$ non identical subsets of
$\mathcal{V}-\{S\}$ such that $D \in \mathcal{V}_i$ for all $1 \leq i \leq
l$. Also assume a product distribution on $x_i,~i \in \mathcal{V}$.
Then
\beq \psi(\mathcal{V}_1,\mathcal{V}_2)+\cdots
+\psi(\mathcal{V}_{l-1},\mathcal{V}_l)+\psi(\mathcal{V}_l,\mathcal{V}_1) \geq
\sum_{i=1}^l \psi(\tilde{\mathcal{V}}_i,\tilde{\mathcal{V}}_i) \eeq where for $k=1,\ldots,l$,
\begin{eqnarray}
\tilde{\mathcal{V}}_{k}&=&\bigcup_{\{i_1,\ldots,i_{k} \} \subseteq
\{1,\ldots,l \}} (\mathcal{V}_{i_1}\cap \cdots \cap \mathcal{V}_{i_{k}} )
\end{eqnarray}
\iffalse
\begin{eqnarray}
\tilde{\mathcal{V}}_1&=& \bigcup_{i_1=1}^l \mathcal{V}_{i_1}\\ \tilde{\mathcal{V}}_2&=&
\bigcup_{\{i_1,i_2 \} \subseteq \{1,\ldots,l \}} ( \mathcal{V}_{i_1}\cap \mathcal{V}_{i_2}
) \\ \nonumber \vdots && \\
\tilde{\mathcal{V}}_{k-1}&=&\bigcup_{\{i_1,\ldots,i_{l-1} \} \subseteq
\{1,\ldots,l \}} (\mathcal{V}_{i_1}\cap \cdots \cap \mathcal{V}_{i_{l-1}} ) \\ \tilde{\mathcal{V}}_l
&=& \mathcal{V}_{1}\cap \cdots \cap \mathcal{V}_{l}
\end{eqnarray}
\fi
or in another words each $\tilde{\mathcal{V}}_j$ is the union of $l \choose
j$ sets such that each set is intersect of $j$ of $\mathcal{V}_i$'s.
\end{lemma}
A special case of this Lemma was recently stated in an independent
work in \cite{brian_submodularity} (Lemma 2) in
the context of erasure networks with only multiple access and no broadcast.

Now we are ready to prove Lemma \ref{lem:trellis_min_cut}.

\begin{proof}(proof of Lemma \ref{lem:trellis_min_cut})
We have
\begin{small}
\beq
H(Y_{\Omega_{\text{unf}}^c}|X_{\Omega_{\text{unf}}^c})=
\sum_{i=1}^{K-1} H(Y_{\mathcal{V}_{i+1}}|X_{\mathcal{V}_i}) = \sum_{i=1}^{K-1}
\psi(\mathcal{V}_i,\mathcal{V}_{i+1})
\eeq
\end{small}
Now look at the sequence of $\mathcal{V}_i$'s. Note that there are total of
$L=2^{|\mathcal{V}|-2}$ possible subsets of $\mathcal{V}$ that contain $D$ but not
$S$. Assume that $\mathcal{V}_s$ is the first set that is revisited. Assume that
it is revisited at step $\mathcal{V}_{s+l}$. Therefore by Lemma \ref{lem:loop}
we have \beq \sum_{i=1}^{l-1} \psi(\mathcal{V}_i,\mathcal{V}_{i+1}) \geq \sum_{i=1}^l
\psi(\tilde{\mathcal{V}}_i, \tilde{\mathcal{V}}_i) \eeq where $\tilde{\mathcal{V}}_i$'s are described
in Lemma \ref{lem:loop}. Now note that any of those $\tilde{\mathcal{V}}_i$
contains $D$ but not $S$ and hence it describes a cut in the original
graph, therefore $\psi(\tilde{\mathcal{V}}_i, \tilde{\mathcal{V}}_i ) \geq \min_{\Omega \in
\Lambda_{D}} H(Y_{\Omega^c}|X_{\Omega^c})$ and hence \beq
\sum_{i=1}^{l-1} \psi(\mathcal{V}_i,\mathcal{V}_{i+1}) \geq l \min_{\Omega \in \Lambda_{D}}
H(Y_{\Omega^c}|X_{\Omega^c})\eeq which means that the value of
that loop is at least length of the loop times the min-cut of the
original graph. Now since in any $L-1$ time frame there is at
least one loop therefore except at most a path of length $L-1$
everything can be replaced with the value of the min-cut in
$\sum_{i=1}^{K-1} \psi(\mathcal{V}_i,\mathcal{V}_{i+1})$. Therefore, \beq \sum_{i=1}^{K-1}
\psi(\mathcal{V}_i,\mathcal{V}_{i+1}) \geq (K-L+1) \min_{\Omega \in \Lambda_{D}}
H(Y_{\Omega^c}|X_{\Omega^c}) \eeq
\end{proof}
%
%In a special case that the distribution of $x_i[t]$'s are all iid and
%uniform over this lemma states our desired result in terms of the
%connection between the min-cut of a network and its time-expanded
%version: \beq (K-\Lambda+1) \min_{\Omega \in \Lambda_{D}}
%\mathrm{rank}(\Gbf_{\Omega,\Omega^c})\log p \leq
%\mathrm{rank}(\Gbf_{\Omega_{\text{unf}},\Omega_{\text{unf}}^c})\log p
%\eeq So what the lemma in this special case says is that the min-cut
%of the time-expanded network (normalized by $K$) can not be much
%smaller than the min-cut of the original network and moreover this gap
%decreases as $K$ increase.

%%\label{sec:Disc}

\vspace{0.1in} \noindent {\bf Acknowledgements}: D. Tse would like to
thank Raymond Yeung for the many discussions on network
coding. S. Diggavi would also like to thank Christina Fragouli for
several enlightening discussions on linear network coding. The
research of D.  Tse and A. Avestimehr are supported by the National
Science Foundation through grant CCR-01-18784 and the ITR grant:"The
3R's of Spectrum Management:Reuse, Reduce and Recycle.". The research
of S. Diggavi is supported in part by the Swiss National Science
Foundation NCCR-MICS center.

\appendix[Proof of Lemma \ref{lem:loop}] \label{app:lem_loop} First we
state a few lemmas some of whose proofs are very straightforward and
hence omitted,
\begin{lemma}
The $\tilde{\mathcal{V}}_i$'s defined in Lemma \ref{lem:loop} satisfy, \beq
\tilde{\mathcal{V}}_l \subseteq \tilde{\mathcal{V}}_{l-1} \subseteq \cdots \subseteq
\tilde{\mathcal{V}}_1\eeq
\end{lemma}

\begin{lemma}\label{lem:ent_equality}
Let $\mathcal{V}_1,\ldots,\mathcal{V}_l$ be $l$ non identical subsets of $\mathcal{V}-\{S\}$ such
that $D \in \mathcal{V}_i$ for all $1 \leq i \leq l$. Also assume that
$\tilde{\mathcal{V}}_1,\ldots,\tilde{\mathcal{V}}_l$ are as defined in lemma
\ref{lem:loop}. Then for any $v \in \mathcal{V}$ we have \beq |\{i|v \in
\mathcal{V}_i\}|=|\{j|v \in \tilde{\mathcal{V}}_j\}|\eeq
\end{lemma}

\begin{proof}
This lemma just states that for each $v \in V$ the number of times
that $v$ appears in $V_i$'s is equal to the number of times that $v$
appears in $\tilde{V}_i$'s. To prove it assume that $v$ appears in
$V_i$'s is $n$. Then clearly \beq v \in \tilde{V}_j, \quad
j=1,\ldots,n \eeq Now for any $j>n$ any element that appears in each
$\tilde{V}_j$ must appear in at least $j$ of $V_i$'s and since $v$
only appears in $n$ of $V_i$'s therefore, \beq v \notin \tilde{V}_j,
\quad j>n \eeq therefore \beq |\{i|v \in V_i\}|=|\{j|v \in
\tilde{V}_j\}|=n \eeq
\end{proof}

\begin{lemma}
Let $\mathcal{V}_1,\ldots,\mathcal{V}_l$ be $l$ non identical subsets of
$\mathcal{V}-\{S\}$ such that $D \in \mathcal{V}_i$ for all $1 \leq i \leq
l$. Also assume a product distribution on $X_i,~i \in
\mathcal{V}$. Then \beq H(X_{\mathcal{V}_1})+\cdots + H(X_{\mathcal{V}_l}) =
H(X_{\tilde{\mathcal{V}}_1}) + \cdots + H(X_{\tilde{\mathcal{V}}_l}) \eeq where
$\tilde{\mathcal{V}}_i$'s are defined in Lemma \ref{lem:loop} and $H(.)$ is just
the binary entropy function.
\end{lemma}

\begin{proof}
For any $v \in V$ define \beq n_v=|\{i|v \in V_i\}| \eeq and \beq
\hat{n}_v=|\{j|v \in \tilde{V}_j\}| \eeq Now since $X_i,~i \in V$ are
independent of each other we have \beq H(X_{V_1})+\cdots + H(X_{V_l})
= \sum_{v \in V} n_v H(X_v) \eeq and \beq H(X_{\tilde{V}_1}) + \cdots
+ H(X_{\tilde{V}_k}) = \sum_{v \in V} \hat{n}_v H(X_v) \eeq By lemma
\ref{lem:ent_equality} we know that $n_v=\hat{n}_v$ for all $v \in V$
hence the lemma is proved.
\end{proof}

The following Lemma is just a straight forward generalization of
submodularity to more than two sets (see also \cite{HKL06}, Theorem 5 where
this result is applied to the entropy  function which is submodular).
\begin{lemma}
\label{lem:k_way_submodularity}
Let $\mathcal{V}_1$, \ldots, $\mathcal{V}_k$ be a collection of sets. Assume that $\xi(\cdot)$
is a submodular function. Then,
\beq \xi(\mathcal{V}_1)+\cdots + \xi(\mathcal{V}_k) \geq
\xi(\tilde{\mathcal{V}}_1) + \cdots + \xi(\tilde{\mathcal{V}}_k)
\eeq
where $\tilde{\mathcal{V}}_i$'s
are defined in Lemma \ref{lem:loop}.

\end{lemma}

Now we are ready to prove Lemma \ref{lem:loop}.
First note that
\begin{scriptsize}
\begin{eqnarray}
\nonumber &&
\psi(\mathcal{V}_1,\mathcal{V}_2)+\cdots+\psi(\mathcal{V}_{l-1},\mathcal{V}_l)+\psi(\mathcal{V}_l,\mathcal{V}_1)=
\\ \nonumber &&
H(Y_{\mathcal{V}_2}|X_{\mathcal{V}_1})+\cdots+H(Y_{\mathcal{V}_l}|X_{\mathcal{V}_{l-1}})+H(Y_{\mathcal{V}_1}|X_{\mathcal{V}_l})=
\\ && \nonumber
H(Y_{\mathcal{V}_2},X_{\mathcal{V}_1})+\cdots+H(Y_{\mathcal{V}_l},X_{\mathcal{V}_{l-1}})+H(Y_{\mathcal{V}_1},X_{\mathcal{V}_l})
- \sum_{i=1}^l H(X_{\mathcal{V}_i}) \end{eqnarray}
\end{scriptsize}
and

\begin{eqnarray} \sum_{i=1}^l \psi(\tilde{\mathcal{V}}_i,\tilde{\mathcal{V}}_i) &=&\sum_{i=1}^l H(Y_{\tilde{\mathcal{V}}_i}|X_{\tilde{\mathcal{V}}_i}) \\
 &=& \sum_{i=1}^l H(Y_{\tilde{\mathcal{V}}_i},X_{\tilde{\mathcal{V}}_i})- \sum_{i=1}^l
 H(X_{\tilde{\mathcal{V}}_i}) \end{eqnarray}

Now define the set \beq \mathcal{W}_i=\{ Y_{\mathcal{V}_i},X_{\mathcal{V}_{i-1}}\}, \quad i=
1,\ldots,l \eeq where $\mathcal{V}_0=\mathcal{V}_l$.  Since by lemma
\ref{lem:ent_equality} we have \beq \sum_{i=1}^l H(X_{\mathcal{V}_i}) =
\sum_{i=1}^l H(X_{\tilde{\mathcal{V}}_i}) \eeq we just need to prove that \beq
\sum_{i=1}^l H(\mathcal{W}_i) \geq \sum_{i=1}^l
H(Y_{\tilde{\mathcal{V}}_i},X_{\tilde{\mathcal{V}}_i}) \eeq Now by since entropy is a
submodular function by Lemma \ref{lem:k_way_submodularity} (k-way
submodularity) we have, \beq \label{eqn:use_k_way} \sum_{i=1}^l H(\mathcal{W}_i)
\geq \sum_{i=1}^l H(\tilde{\mathcal{W}}_i)\eeq where \beq
\tilde{\mathcal{W}}_r=\bigcup_{\{i_1,\ldots,i_r\} \subseteq \{1,\ldots,l \}}
(\mathcal{W}_{i_1} \cap \cdots \cap \mathcal{W}_{i_r} ), \quad r=1,\ldots,l \eeq Now for
any $r$ ($1 \leq r \leq l$) we have

\begin{eqnarray*}
\displaystyle
\tilde{\mathcal{W}}_r&=&\bigcup_{\{i_1,\ldots,i_r\} \subseteq \{1,\ldots,l \}}
(\mathcal{W}_{i_1} \cap \cdots \cap \mathcal{W}_{i_r}) \\ &=& \bigcup_{\{i_1,\ldots,i_r\}
\subseteq \{1,\ldots,l \}} (\{ Y_{\mathcal{V}_{i_1}},X_{\mathcal{V}_{i_1-1}} \} \cap
\cdots \cap \{ Y_{\mathcal{V}_{i_r}}X_{\mathcal{V}_{i_r-1}} \}) \\ &=&
\bigcup_{\{i_1,\ldots,i_r\} \subseteq \{1,\ldots,l \}} (\{ Y_{\mathcal{V}_{i_1}
\cap \cdots \cap \mathcal{V}_{i_r} },X_{\mathcal{V}_{(i_1-1)} \cap \cdots \cap
X_{\mathcal{V}_{(i_r-1)}}} \}) \\ \nonumber &=& \left \{
Y_{\bigcup_{\{i_1,\ldots,i_r\} } (\mathcal{V}_{i_1} \cap \cdots \cap \mathcal{V}_{i_r})
},X_{\bigcup_{\{i_1,\ldots,i_r\} }(\mathcal{V}_{(i_1-1)} \cap \cdots \cap
\mathcal{V}_{(i_r-1)})} \right \} \\ &=& \{Y_{\tilde{\mathcal{V}}_r},X_{\tilde{\mathcal{V}}_r} \}
\end{eqnarray*} Therefore by equation (\ref{eqn:use_k_way}) we have,
\begin{eqnarray} \sum_{i=1}^l H(\mathcal{W}_i)
& \geq & \sum_{i=1}^l H(\tilde{\mathcal{W}}_i) \\
&=&\sum_{i=1}^l H(Y_{\tilde{\mathcal{V}}_i},X_{\tilde{\mathcal{V}}_i})
\end{eqnarray}

Hence the Lemma is proved.

\end{document}